\documentclass[aps,prl,twocolumn,showpacs,floatfix,amsmath]{revtex4}
\usepackage{graphicx}

\newcommand{\braket}[1]{\langle #1 \rangle}
\begin{document}

\title{Three-dimensional Roton-Excitations and Supersolid formation in Rydberg-excited Bose-Einstein Condensates}
\author{N. Henkel, R. Nath and T. Pohl}
\affiliation{Max Planck Institute for the Physics of Complex Systems, N\"othnitzer Stra\ss e 38, 01187 Dresden, Germany}

\begin{abstract}
We study the behavior of a Bose-Einstein condensate in which atoms are weakly coupled
to a highly excited Rydberg state. Since the latter have very strong van der Waals interactions,
this coupling induces effective, nonlocal interactions between the dressed groundstate atoms,
which, opposed to dipolar interactions, are isotropically repulsive. Yet, one finds partial attraction
in momentum space, giving rise to a roton-maxon excitation spectrum and a transition to a super
solid state in three-dimensional condensates. A detailed analysis of decoherence and loss
mechanisms, suggests that these phenomena are observable with current experimental capabilities.
\end{abstract}


\maketitle
Since introduced by Landau in a series of seminal articles \cite{landau}, the notion of a roton
minimum in the dispersion of a quantum liquid has been pivotal to understanding
superfluidity in helium. This later led to the prediction of a peculiar solid state
upon softening of the roton excitation energy \cite{softening}, simultaneously possessing
crystalline and superfluid properties. In such a supersolid \cite{suso1}, the particles that
must supply the rigidity to form a crystal, at the same time provide for superfluid nonviscous
flow. Forty years after its conjecture, this apparent contradiction continues to attract
theoretical interest and has ushered in an intense search for experimental evidence in
solid $^4$He, whose interpretations are currently under active debate \cite{suso2,bps06}.

\begin{figure}[t!]
\begin{center}
\resizebox{0.99\columnwidth}{!}{\includegraphics{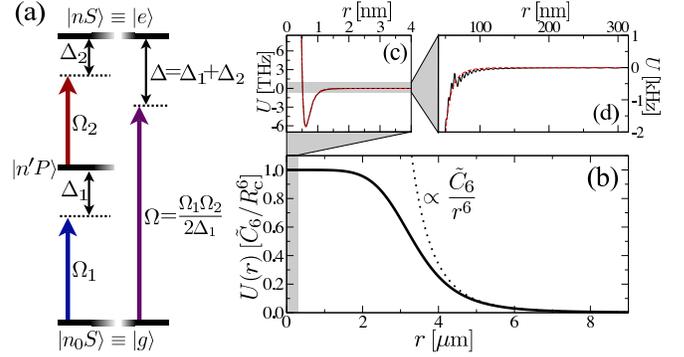}}
\caption{\label{fig1}(a) Schematics of the considered three-level atom, illustrating the laser coupling between the atomic groundstate $|n_0S\rangle$ and the Rydberg state $|nS\rangle$. For $\Delta_1\gg \Omega_1$, the system reduces to an effective two-level atom, with the states $|g\rangle\equiv |n_0S\rangle$ and $|e\rangle\equiv |nS\rangle$ coupled with a two-photon Rabi frequency $\Omega$ and detuning $\Delta$. (b) Effective potential resulting from the off-resonant coupling to the strongly interacting Rydberg states for $n=60$ and $\Delta=50$MHz. Panels (c) and (d) provide an enlarged view of the potential showing the  contributions from both groundstate-Rydberg atom and groundstate-groundstate atom interactions (solid line) as well as the sole contribution from the latter (dashed line).}
\end{center}
\end{figure}

Here, we demonstrate how three-dimensional roton excitations can be realized
in atomic Bose-Einstein condensates (BECs), thereby introducing an alternative
system to study supersolidity. The supersolid phase transition is shown to arise from
effective interactions, realized through off-resonant optical coupling \cite{ssz00,mls09,gmb09}
to highly excited Rydberg states.
Owing to the strong increase of atomic interactions with their principal quantum number
$n$, {\emph{resonantly}} excited Rydberg gases have proved to be an ideal platform to
study strong interactions in many-body systems \cite{rydberg} on short $\mu$s
time scales. The present approach $-$ based on {\emph{off-resonant}}  two-photon excitation
[see Fig.\ref{fig1}a] of Bose-condensed alkaline atoms $-$ permits to utilize the strong Rydberg
interactions over much longer times of $\sim100$ms. In particular, we consider coupling to $nS$
Rydberg states with vanishing orbital angular momentum, which, as opposed to dipole-dipole
interactions, gives rise to isotropically repulsive interaction potentials for the groundstate
atoms, and, thus, ensures stability of the condensate.

The system is described as a gas of $N$ atoms with mass $M$ at positions ${\bf r}_i$, each possessing a ground state $|g_i\rangle$ and
an excited $nS$ Rydberg state, denoted by  $|e_i\rangle$. The two states are optically coupled with a two-photon Rabi frequency
$\Omega$ and detuning $\Delta$ (see Fig.\ref{fig1}a). Defining corresponding transition and projection
operators $\hat{\sigma}_{\alpha\beta}^{(i)}=|\alpha_i\rangle\langle\beta_i|$ ($\alpha,\beta={\rm e,g}$), the resulting $N$-particle
interaction can be written as
\begin{eqnarray}\label{ham}
\hat{H}_{I} &=&\sum_{i<j}V_{\rm ee}({\bf r}_{ij})\hat{\sigma}_{\rm ee}^{(i)}\hat{\sigma}_{\rm ee}^{(j)} - \hbar\Delta\sum_i\hat{\sigma}_{\rm ee}^{(i)} + \hat{H}_{L},\nonumber
\end{eqnarray}
where $\hat{H}_{L} =\tfrac{\hbar\Omega}{2}\sum_i\hat{\sigma}_{\rm eg}^{(i)}+\hat{\sigma}_{\rm ge}^{(i)}$ describes the laser coupling and $V_{\rm ee}({\bf r}_{ij})=C_6/r_{ij}^6>0$ denotes the van der Waals (vdW) interaction between two Rydberg atoms at a distance ${\bf r}_{ij}={\bf r}_{i}-{\bf r}_{j}$.
Because of the strong $C_6\sim n^{11}$ scaling of the vdW coefficient, such Rydberg-Rydberg atom interactions are orders of magnitude larger than
those of groundstate atoms. We are interested in the potential surface $W_{G}({\bf r}_1,...,{\bf r}_N)$ that asymptotically connects to the many-body groundstate
$|G\rangle=\bigotimes_k|g_k\rangle$, under the condition of far-off resonant driving $\Omega/|\Delta|\ll 1$ and for $\Delta<0$. Under the latter
condition the $N$-body potential energy of the many-body groundstate $|G\rangle$ is separated from the excited states by at least $\Delta$ \cite{pdl09},
justifying the use of a Born-Oppenheimer (BO) treatment for the atomic dynamics on the potential surface $W_{G}$.
The BO surface $W_{G}$ can be determined from a many-body perturbation expansion in the small parameter $\Omega/|\Delta|\ll1$, which up to fourth order only involves coupling to the singly and doubly excited many-body states $|E_i\rangle=|e_i\rangle\bigotimes_{k\neq i}|g_k\rangle$ and $|E_{ij}\rangle=|e_ie_j\rangle\bigotimes_{k\neq i,j}|g_k\rangle$. Explicitly, we obtain
\begin{eqnarray}
W_G &=& 2\sum_{i\neq j}\tfrac{|\braket{G|\hat H_{\rm L}|E_i}\braket{E_i|\hat H_{\rm L}|E_{ij}}|^2}{ \hbar^2\Delta^2(2\hbar\Delta-V_{\rm ee}({\bf r}_{ij}))}+\sum_{i}\tfrac{|\braket{G|\hat H_{\rm L}|E_i}|^2}{\hbar\Delta} \nonumber\\
 &=&  \sum_{i \neq  j}\tfrac{\hbar ^4\Omega^4}{8\hbar^2\Delta^2(2\hbar\Delta-V_{\rm ee}({\bf r}_{ij}))} + \tfrac{\hbar^2\Omega^2N}{4\hbar\Delta} \nonumber\\
 &=&\tfrac{N\hbar^2\Omega^2}{4\hbar\Delta}+\tfrac{N(N-1)\hbar^4\Omega^4}{16\hbar^3\Delta^3}+\sum_{i<j}\tfrac{\tilde{C}_6}{r_{ij}^6+R_{\rm c}^6}
\end{eqnarray}
Omitting constant terms, the leading-order, $N$-body BO surface is given by a sum of binary effective potentials $U(r_{ij})=\tilde{C}_6/(r_{ij}^6+R_{\rm c}^6)$, exemplarily shown in Fig.\ref{fig1}b. \\
The peculiar shape of $U$ is easily understood within a simple two-atom picture: For far-distant atomic pairs ($r_{ij}\gg R_{\rm c}$), a small fraction $(\Omega/2\Delta)^2$ is independently admixed to each groundstate atom, such that $U(r_{ij})$ is of vdW-type with an effective coefficient $\tilde{C}_6=(\Omega/2\Delta)^4C_6$.
At smaller distances, however, the interaction shift $V_{\rm ee}({\bf r}_{ij})$ renders dressing to the doubly excited $|E_{ij}\rangle$ states ineffective, such that the effective potential approaches a constant value below a critical distance $R_{\rm c}=(C_6/2\hbar |\Delta|)^{1/6}$. Typically, $R_{\rm c}$ can take on rather large values of a few $\mu$m.

\begin{figure}[t!]
\begin{center}
\resizebox{0.99\columnwidth}{!}{\includegraphics{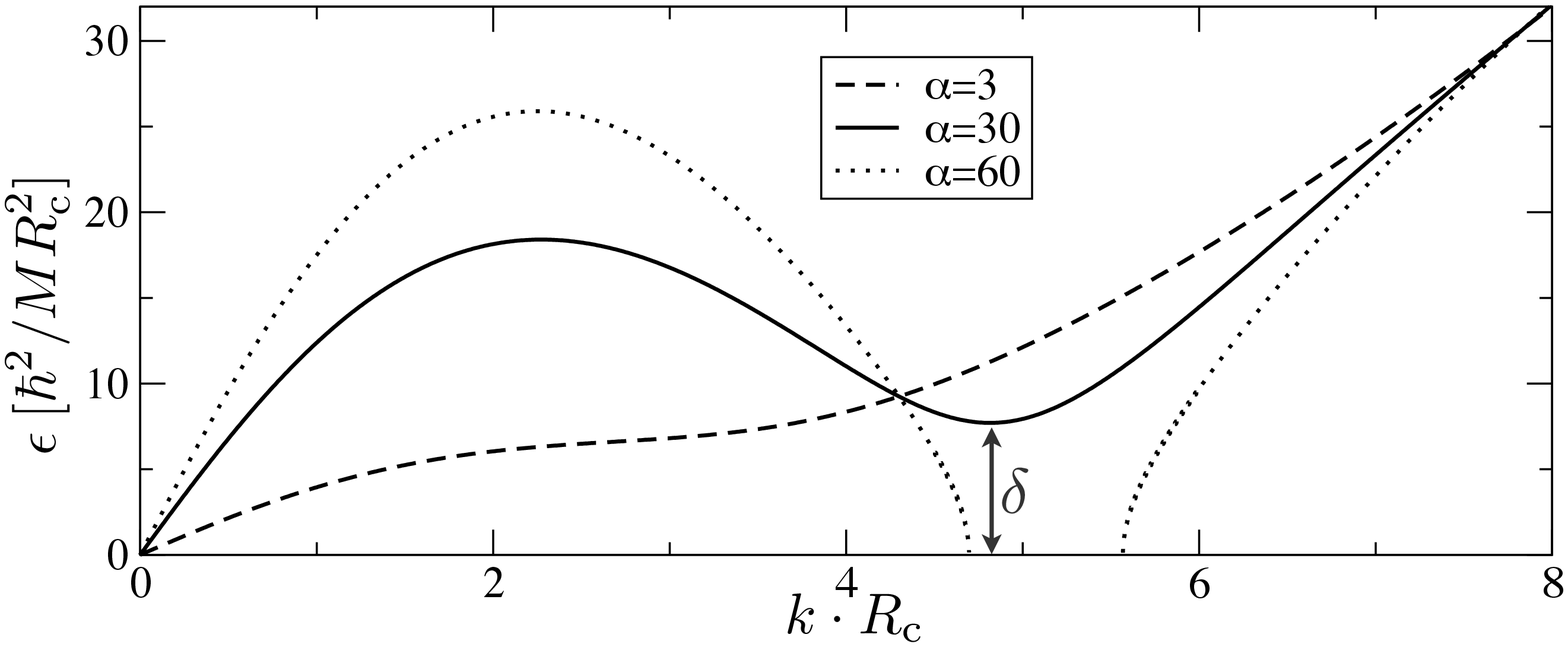}}
\caption{\label{fig2}Dispersion relation $\epsilon(k)$ for different values of the interaction parameter $\alpha$. The arrow indicates the roton gap $\delta$.}
\end{center}
\end{figure}

\begin{figure}[b!]
\begin{center}
\resizebox{0.95\columnwidth}{!}{\includegraphics{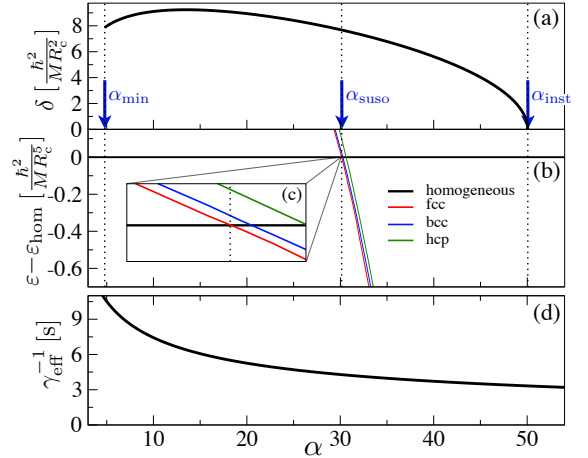}}
\caption{\label{fig3}(a) Roton gap $\delta$ as a function of the interaction parameter $\alpha$. (b) Energy density $\varepsilon=\tfrac{\rho_0}{2N}\langle\Psi| \tfrac{-\hbar^2\nabla^2}{2M}+\hat{H}|\Psi\rangle$  for different crystal symmetries relative to the energy density $\varepsilon_{\rm hom}=\pi^2\alpha/3$ of a homogeneous BEC. Panel (c) provides an enlarged view around the transition point. The effective Rydberg state lifetime for excitation of $^{87}$Rb to $n=60$($C_6=9.7\cdot10^{20}\rm{a.u.}$ \cite{sk05}\,) with $\Delta=50$MHz and $\rho_0=10^{14}$cm$^{-3}$ is shown in (d).}
\end{center}
\end{figure}

For simplicity, the above discussion has ignored interactions between pairs of groundstate atoms as well as groundstate-Rydberg atom interactions. Both, however, can  be included in the calculations of the effective potential (see Fig.\ref{fig1}c). Although being comparably strong, the range of the corresponding contributions is considerably smaller than $R_{\rm c}$. Upon avoiding photo-excitation of Rydberg-molecular resonances \cite{bbn09} these additional potentials may, thus, be described in terms of an s-wave scattering pseudopotential, with a scattering length $a$. On the other hand, the large value of $R_c$ prevents such a simplified treatment for $U(r)$. However, as long as $\tilde{C}_6\ll \hbar^2R_{\rm c}^4/M$ its effect can be described within a first Born approximation \cite{firstB} \footnote{For the parameters of Fig.\ref{fig3}c $\tilde{C}_6M/\hbar^2R_{\rm c}^4<10^{-2}$.}.
With these simplifications and in the zero-temperature limit one arrives at the following nonlocal, nonlinear Schr\"odinger equation
\begin{equation}
i\hbar\frac{\partial \Psi}{\partial t}=
\left [-\frac{\hbar^2\nabla^2}{2M}+g|\Psi|^2+\int d{\bf r}^{\prime} U({\bf r}-{\bf r}^{\prime}) |\Psi({\bf r}^{\prime})|^2
\right ]\Psi,
\label{GPE}
\end{equation}
for the condensate wavefunction $\Psi(\bf r)$. The contact interaction term $\propto g=4\pi\hbar^2 a/M$ may be tuned to very small values.  For clarity, it will, hence, be omitted  in the following and briefly discussed at the end of this work.

For small $\tilde{C}_{6}$ the BEC ground state corresponds to a homogeneous superfluid with density $\rho_0$. Its elementary
excitations with wave number ${\bf k}$ and corresponding energy $\epsilon$ are calculated from the corresponding Bogoliubov-de Gennes equations, which yield
$\epsilon(k)^2 = \tfrac{\hbar^2k^2}{2M} [\tfrac{\hbar^2k^2}{2M} +2 \rho_0 \tilde{U}(k)]$, where $\tilde{U}(k)$ is the Fourier transform of the interaction potential $U(r)$.
Upon appropriate length and energy scaling (see Fig.\ref{fig2}), the dispersion is determined by a single dimensionless parameter
$\alpha=\rho_0M\tilde{C}_6/\hbar^2R_{\rm c}$, parametrizing the interaction strength.
Asymptotically the dispersion relation $\epsilon(k)$ has phonon and free particle character, at small and large $k$,
respectively. However, due to the inner potential plateau at $r<R_{\rm c}$, the momentum space potential $\tilde{U}$ has negative attractive contributions around
$k=k_{\rm rot}\sim2\pi/R_{\rm c}$, such that the spectrum develops a roton minimum at $k=k_{\rm rot}$ (see Fig.\ref{fig2})
for sufficiently large $\alpha>\alpha_{\rm rot}\approx4.8$. The corresponding roton gap $\delta$
decreases with increasing $\alpha$ and ultimately vanishes at $\alpha_{\rm inst}\approx50.1$ (see Fig.\ref{fig3}a), marking the onset of
a roton instability, at which density modulations may develop without energy cost.
\begin{figure}[t!]
\begin{center}
\resizebox{0.90 \columnwidth}{!}{\includegraphics{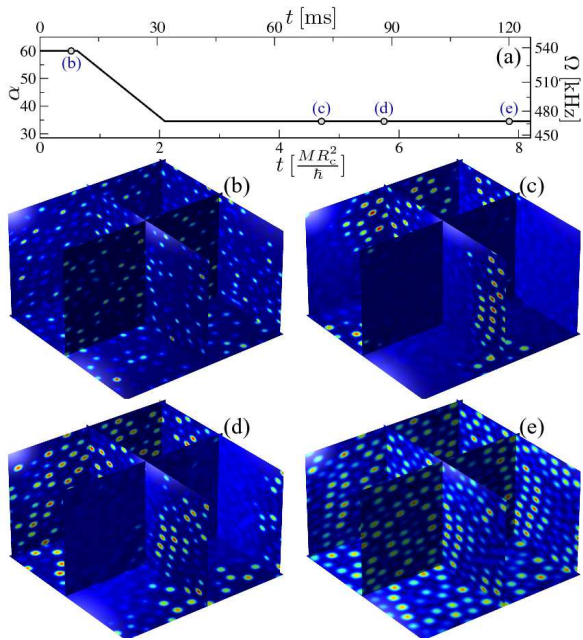}}
\caption{\label{fig4}Snapshots of the BEC dynamics for a time-varying interaction parameter $\alpha(t)$ shown in (a). Panels (b)-(e) show the density along orthogonal slices through the simulation box at times indicated in (a). The upper and right axes in (a) show the actual time and Rabi frequency for a $^{87}$Rb BEC with $n=60$, $\Delta=50$MHz and $\rho_0=2\cdot10^{14}$cm$^{-3}$.}
\end{center}
\end{figure}
In ultracold gases, a similar behavior was found in tightly confined, pancake-shaped dipolar
BECs \cite{gri05,lms09}. While it is the partially attractive nature of the anisotropic interactions that
generates the 2$D$ roton excitations in this case \cite{ssl03}, it also leads to collapse
in the post-roton-instability phase, thus, precluding formation of supersolid density
modulations  \cite{kc07,dkm07} in higher than one dimension \cite{gdk02}.
The present interactions, on the other hand, are entirely repulsive, which may indeed provide for
stable regular density modulations.

Addressing this question, requires us to go beyond the linear analysis discussed above. We performed numerical simulations of Eq.(\ref{GPE}) on a three-dimensional grid of $\sim10^7$ grid
points with periodic boundary conditions.
Because of the nonlinear character of the Gross Pitaevskii equation a standard imaginary-time integration
may generally not converge to the true groundstate of the system. As also found for lattice-confined
dipolar bosons \cite{mtl07}, there is a large number of stationary states, that correspond to local minima
of the total energy. As a consequence, the system generally approaches a glassy state with short-range
ordered density modulations, when starting from a homogenous initial state (see below). We, thus, used variational calculations, based on periodically arranged Gaussians with varying width and lattice
constant, to provide the proper initial wavefunction for a subsequent imaginary time-evolution according
to Eq.(\ref{GPE}).

Some of the obtained energies are shown in Fig.\ref{fig3}b. For small values of $\alpha$ the BEC
groundstate is a homogenous superfluid. At a critical value of $\alpha_{\rm suso}\approx30.1$, one finds
a transition to a stable supersolid state. This first-order transition precedes the roton-instability \cite{pr94}
and takes place at a finite roton gap of $\delta_{\rm suso}\approx 0.66 \tfrac{\hbar^2k_{\rm rot}^2}{2M}$.
The existence of several competing states with similar energies but different crystal symmetries (see Fig.\ref{fig3}b),
may generally complicate the experimental preparation of ordered states. In this respect, the dynamical
tunability of the interaction strength $\tilde{C}_6$ via changing the laser intensity can serve a useful tool
to steer the BEC evolution.

In order to demonstrate this point, we also studied the time evolution, starting from a homogenous BEC.
As a specific example, we discuss the BEC dynamics for a simple time-dependence of
 $\alpha$, shown in  Fig.\ref{fig4}a. The calculation starts from a homogenous condensate with small random phase noise
and uses a complex-time integration with a small imaginary contribution \cite{slv02}. The instantaneous
increase of $\alpha$ at time $t=0$ from $\alpha=0$ to $60$ induces the roton instability. This sudden parameter
quench, however, causes relaxation towards a short-range ordered, "glassy" \cite{bps06} state (Fig.\ref{fig4}b), as discussed above.
As $\alpha$ is decreased close to the phase transition some of the structures vanish entirely, leading to a mixed phase in which extended
superfluid fractions of nearly constant density coexist with density-modulated domains (Fig.\ref{fig4}c). The latter increase in time (Fig.\ref{fig4}d), and
ultimately merge to form sizable "crystallites" of regular density modulations (Fig.\ref{fig4}e).

\begin{figure}[b!]
\begin{center}
\resizebox{0.95\columnwidth}{!}{\includegraphics{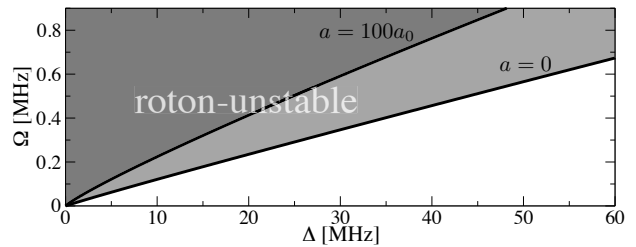}}
\caption{\label{fig5}Required laser parameters to induce the roton instability (solid lines) for $n=60$ and $\rho_0=10^{14}$cm$^{-3}$.
The upper curve also accounts for s-wave scattering with $a=100a_0$.}
\end{center}
\end{figure}

Turning to a discussion of the experimental feasibility, we consider a particular example of coupling to $60S$
Rydberg states in a $^{87}$Rb condensate, for which Rydberg excitation has recently been demonstrated
\cite{hrb08}. Fig.\ref{fig5} shows the corresponding "phase diagram" for a typical density of $10^{14}$cm$^{-3}$,
also including a finite s-wave scattering length $a>0$. The latter only leads to some increase of the critical Rabi
frequencies for inducing the roton instability.
Importantly, the transition to a supersolid can be realized with Rabi frequencies of a few hundred kHz, and the condition $|\Delta|\gg \Omega$ can
be well fulfilled deep in the roton-instability regime. Yet, the detuning is sufficiently small to avoid excitation of nearby Rydberg states
and near-resonant dipole coupling to adjacent pair states for distances $\gtrsim R_{\rm c}$. For typical values of $R_{\rm c}$, the
number of atoms within a single density peak is on the order of
$10^3$, which justifies the applied meanfield description in terms of Eq.(\ref{GPE}).

Major limitations for the stability of Rydberg gases generally stem  from the
finite lifetime of the involved excited states and from auto-ionization of close Rydberg atom pairs, initiated
by near-resonant dipole-dipole couplings to energetically close pair states \cite{ths08}.
In the present case, the latter are suppressed by a factor $(\tfrac{\Omega}{2\Delta})^4\tfrac{r_{ij}^{12}}{(r_{ij}^6+R_{\rm c}^6)^2}$ due to the interaction blockade of doubly excited states, and, thus, limited to negligible small values of $30$mHz for the parameters range of Fig.\ref{fig3}d.
Rydberg state decay, due to spontaneous emission and black-body radiation, is more significant with typical rates on the
order of $\gamma\sim10$ms$^{-1}$. However, since the Rydberg state is only weakly populated by the far
off-resonant coupling to the groundstate, the resulting effective decay rate $\gamma_{\rm eff}=(\Omega/2\Delta)^2\gamma$
can be decreased to much smaller values. As shown in Fig.\ref{fig3}d, for typical atomic densities and laser parameters the
effective lifetime is as large as several seconds, over the entire range of relevant interaction parameters.
A second loss mechanism arises from spontaneous decay of the intermediate $P$ state (with rate $\gamma^{\prime}$),
used to drive the $5S\rightarrow n^{\prime}P\rightarrow nS$ two-photon transition (see Fig.\ref{fig1}a). Again,
the corresponding effective  decay rate $\gamma^{\prime}_{\rm eff}=(\Omega_1/2\Delta_1)^2\gamma^{\prime}$ is greatly suppressed
for small ratios $\Omega_1/\Delta_1$, which, however, also reduces the two-photon Rabi frequency $\Omega=\Omega_1\Omega_2/2\Delta_1$.
 With $\Omega_2$ of a few hundred MHz and $\Omega_1/\Delta_1=10^{-3}$ one
realizes a two-photon Rabi frequency of several hundred kHz, as considered in the example of Fig.\ref{fig5}. Such large
Rydberg Rabi frequencies ($\Omega_2$), could be achieved by choosing a $6P$ state for the two-photon transition, allowing to
drive the Rydberg transition with a strong $780$nm-laser. At the same time this gives a longer intermediate-state lifetime
$\gamma^{\prime-1}=121$ns \cite{gao04}, and with $\Omega_1/\Delta_1=10^{-3}$ yields long effective lifetimes of
$\gamma^{\prime-1}_{\rm eff}=0.5$s. Together with the small value of $\gamma_{\rm eff}$ this ensures sufficiently low loss rates
to observe the phenomena discussed in this work (cf. Fig.\ref{fig4}).

In summary, we have shown that off-resonant Rydberg excitation of ultracold atoms provides a promising route for the preparation of
three-dimensional supersolid phases in Bose-Einstein condensates. The unambiguous realization of this elusive state opens up
a range of new studies, from investigations of thermal effects to supersolid formation in finite systems and their dynamical response
to perturbations, such as trap rotations.

Generally, the described setting offers new avenues for the realization of complex nonlocal media, where the sign, shape
and strength of interactions can be tuned through proper choice of the addressed Rydberg state and the applied laser
parameters, and even permits local spatial control of nonlocal interactions via tightly focused beams.

We thank T. Pfau, N. V. Prokof'ev, L. Santos, I. Lesanovsky, G. Pupillo, V. Pai and S. Skupin  for helpful discussions, and W. Li for assistance in
calculating groundstate-Rydberg atom interactions.

\end{document}